# Discretized dynamics of exchange spin wave bulk and edge modes in honeycomb nanoribbons with armchair edge boundaries

D. Ghader[(1)] and A. Khater[(2,3)]

[1] College of Engineering and Technology, American University of the Middle East, Kuwait

[2] Department of Theoretical Physics, Institute of Physics, Jan Dlugosz University in Czestochowa, Am. Armii Krajowej 13/15, Czestochowa, Poland

[3] Department of Physics, University du Maine, 72085 Le Mans, France

**We develop a field theory to study the dynamics of long wavelength exchange spin wave excitations on honeycomb nanoribbons characterized by armchair edge boundaries and the Néel antiferromagnetic ordering state. Appropriate boundary conditions are established by requiring that the bulk and edge spins precess with the same frequency for any given spin wave eigenmode in these systems. A set of characteristic boundary equations, common for bulk and edge spin wave modes, are hence derived. The equations of motion for the spin dynamics are then solved to determine the propagating and evanescent exchange spin wave modes. We prove in general that the bulk spin wave dynamics is discretized due to the finite width of the nanoribbon. For an isotropic magnetic nanoribbon, the Dirac cone is reduced to a single linear dispersion curve due to this discretization. The number and wavelengths of allowed bulk modes for isotropic and anisotropic nanoribbons are determined from the derived characteristic boundary equations. As witnessed by our numerical results for different examples it is shown that the characteristics of these modes depend on the width of the nanoribbon and its antiferromagnetic anisotropy. Further, anisotropic nanoribbons, even those with the slightest anisotropy, present evanescent modes with non-linear dispersion relations. The spatial variation of the amplitudes of the evanescent exchange spin waves across the finite widths of the nanoribbons, is found to be strongly dependent on the system magnetic anisotropy and its width. The developed theoretical approach is general and can be applied for nanoribbons with all types of boundary edges.**

## *Introduction*

The study of magnetic excitations on the boundaries of a finite magnetic system is not new, as it dates to the early nineteen sixties with the pioneering work of Damon and Eshbach [1], and their prediction of the DE dipolar spin wave mode on the surfaces of a ferromagnetic slab. The DE mode was derived using the classical field theory for dipolar spin waves, conjugated with the Maxwell boundary conditions on both surfaces of the ferromagnetic thick film. Similar field theory approaches, with lattice spins treated as classical spins, were then applied to study the surface dipolar spin waves in antiferromagnets [2-7]. More recently, the technological advances in fabricating precision thin and ultrathin magnetic films, combined with very refined experimentation, have stimulated important theoretical research on the dipole-exchange surface spin waves for nano-structured systems.

The classical field theory in the continuum limit was also intensively developed and widely applied to study the bulk and surface exchange magnetic excitations for bounded magnetic systems [8-11]. The exchange interaction is treated using the semi-classical Heisenberg Hamiltonian, and several equivalent exchange-boundary conditions are derived and solved consistently with the bulk equations of motion to determine the allowed bulk and surface spin wave excitations. A particularly efficient exchange boundary condition is derived from the requirement that the equations of motion of a surface spin should be the same as those of a spin in the bulk. The theoretical studies on thin and ultrathin magnetic films demonstrated fundamental effects for both bulk and surface spin wave modes, induced by the structured film surface and its thicknesses. The semi-classical Heisenberg model approach proved very efficient in the linear spin wave theory when the quantum effects are negligible, and was further developed and applied to study spin wave excitations and their scattering in multilayers, surface structures, and nanojunctions [12-21].

The recent emergence of Dirac materials [22] turned the attention towards edge magnetic excitations as an interesting novel fundamental phenomenon with important potentials for technological applications [23-26]. Edge spin waves (or magnons) in 2D honeycomb bounded magnetic system with various edge types have been intensively investigated using quantum approaches, notably the Holstein-Primakov formalism [27-33]. Like surface spin wave excitations, edge spin waves are found to depend on the edge structure.

Despite its leading role in determining the surface and interface spin waves [1-21], the classical field spin wave theory has not yet been systematically developed for edge spin waves in finite 2D honeycomb lattices. In the present paper, the theory is developed and applied to calculate and analyze the long wavelength propagating (bulk) and evanescent (edge) exchange spin wave modes on 2D honeycomb nanoribbons, in the Néel antiferromagnetic ordering state, with armchair edge boundaries.

Using the semi-classical exchange Heisenberg Hamiltonian, we derive equations of motion equivalent to the Harper equations previously derived using the Holstein-Primakov formalism [28-31, 33]. Appropriate exchange boundary conditions are derived by requiring that the equations of motion of an edge spin are the same as those of a bulk spin. This is a necessary and sufficient condition to ensure that edge and bulk spins oscillate at the same frequency for a given eigenmode [8-11]. The boundary conditions yield unified characteristic equations for bulk and edge spin waves. Solving the boundary characteristic equations consistently with the equations of motion, yields the bulk and edge spin waves. Note that the boundary conditions, solved simultaneously on both edges of the nanoribbon, pose constraints on the wavelength of the allowed bulk spin waves in the finite direction of the nanoribbon. The bulk spin wave spectrum is proved to be finite and discrete. The existence of evanescent modes in nanoribbons with armchair edge boundaries is found to be directly related to the magnetic anisotropy. The important effects of magnetic exchange anisotropy and finite nanoribbon width on the characteristics of bulk and edge spin wave modes are analyzed in details.

## *Field theory formulation*

A schematic representation of the nanoribbon is presented in figure (1). The nanoribbon is considered infinite in the y-direction, finite in the x-direction, and with armchair boundaries on its edges. As shown in the figure, the nanoribbon terminates at the edges $x = \pm d = \pm n \times a$ where $a$ is the honeycomb lattice constant and $n$ is an integer. In the Néel ordering state, the spins on A and B sublattices are conventionally assumed to be aligned parallel and antiparallel to the z-axis respectively.

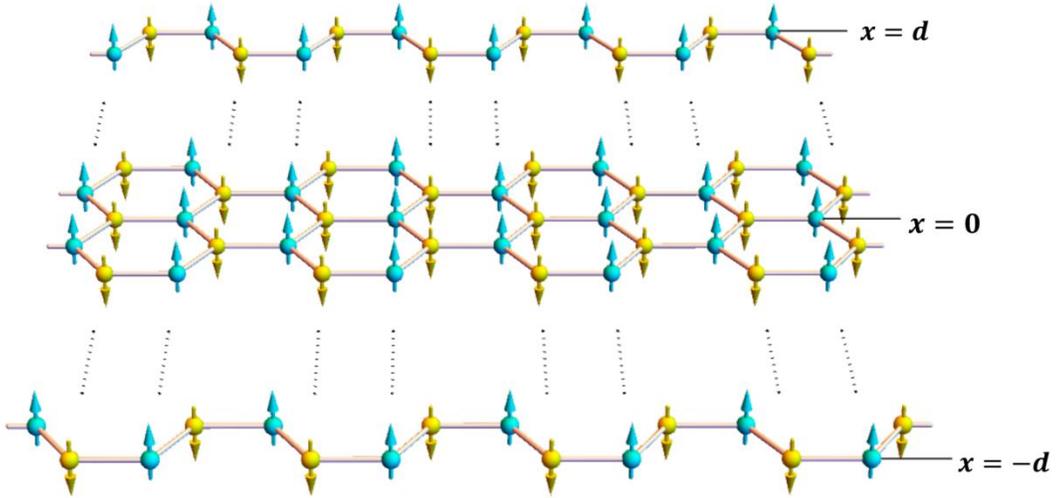

**Figure 1:** Schematic representation of a honeycomb nanoribbon with armchair edge boundaries. A and B sublattice spins are conventionally assumed to be respectively aligned parallel and antiparallel to the z-axis in the Néel magnetic ordering state.

We consider a semi-classical Heisenberg Hamiltonian with anisotropic exchange interactions between nearest neighbors

$$\mathcal{H} = J \sum_{\langle \vec{r}, \vec{\delta} \rangle} \left[ \vec{S}_\parallel(\vec{r}, t) \cdot \vec{S}_\parallel(\vec{r} + \vec{\delta}, t) + \gamma S_z(\vec{r}, t) S_z(\vec{r} + \vec{\delta}, t) \right]$$

where $t$ is time, $J$ is the exchange constant, $\vec{r} = x\,\hat{x} + y\,\hat{y}$ is the position vector of a site on the honeycomb lattice, and $\vec{\delta}$ is the position vector of a nearest neighbor. The vector $\vec{S}_\parallel = S_x \hat{x} + S_y \hat{y}$ is the spin component in the plane of the honeycomb lattice. $\gamma > 1$ is the bulk anisotropy parameter.

In the classical field approach, the magnetizations for A and B sublattices are proportional to the spin vectors and satisfy the Bloch (or Landau-Lifshitz) equations of motion [1-21]

$$\partial_t \vec{M}^A = \lambda \vec{M}^A \times \vec{H}^A \quad (1a)$$
$$\partial_t \vec{M}^B = \lambda \vec{M}^B \times \vec{H}^B \quad (1b)$$

with $\partial_t = \partial/\partial t$, $\lambda$ is the gyromagnetic ratio, and $\vec{H}^A$ and $\vec{H}^B$ denote the effective exchange fields acting on the magnetizations $\vec{M}^A$ and $\vec{M}^B$ respectively; these fields are determined from the Heisenberg Hamiltonian as

$$\vec{H}^A = J \sum_{\vec{\delta}} \left[ \vec{M}^B_{\parallel}(\vec{r}+\vec{\delta},t) + \gamma M^B_z(\vec{r}+\vec{\delta},t) \right]$$

$$\vec{H}^B = J \sum_{\vec{\delta}} \left[ \vec{M}^A_{\parallel}(\vec{r}+\vec{\delta},t) + \gamma M^A_z(\vec{r}+\vec{\delta},t) \right]$$

In the continuum limit and for spin waves with long wavelengths, the magnetizations can be written [8] as

$$\vec{M}^{A/B}(\vec{r}+\vec{\delta},t) = e^{\vec{\delta}\cdot\vec{\nabla}} \vec{M}^{A/B}(\vec{r},t) \approx \left[ 1 + \vec{\delta}\cdot\vec{\nabla} + \frac{1}{2}(\vec{\delta}\cdot\vec{\nabla})^2 \right] \vec{M}^{A/B}(\vec{r},t) \quad (2)$$

Substituting equation (2) in the effective exchange fields, and summing over all nearest neighbors for a bulk site, yields

$$\vec{H}^A = J\left(3 + \frac{a^2}{4}\Delta\right) \vec{M}^B_{\parallel} - 3\gamma J M \hat{z} \quad (3a)$$
$$\vec{H}^B = J\left(3 + \frac{a^2}{4}\Delta\right) \vec{M}^A_{\parallel} + 3\gamma J M \hat{z} \quad (3b)$$

with $M = M^A_z = -M^B_z$ and $\Delta = \frac{\partial^2}{\partial x^2} + \frac{\partial^2}{\partial y^2} = \partial_x^2 + \partial_y^2$. Equations (3) represent the bulk effective exchange fields which are different from the edge fields $\vec{H}_e^{A/B}$ because of the missing nearest neighbors.

Substituting equations (3) in the Bloch equations of motion yields four scalar differential equations for the magnetizations components as follows

$$\partial_t M^A_x = -3\gamma\lambda J M\, M^A_y - \lambda J M \left(3 + \frac{a^2}{4}\Delta\right) M^B_y \quad (4a)$$

$$\partial_t M^A_y = 3\gamma\lambda J M\, M^A_x + \lambda J M \left(3 + \frac{a^2}{4}\Delta\right) M^B_x \quad (4b)$$

$$\partial_t M^B_x = 3\gamma\lambda J M\, M^B_y + \lambda J M \left(3 + \frac{a^2}{4}\Delta\right) M^A_y \quad (4c)$$

$$\partial_t M^B_y = -3\gamma\lambda J M\, M^B_x - \lambda J M \left(3 + \frac{a^2}{4}\Delta\right) M^A_x \quad (4d)$$

Deriving equation (4a) with respect to time and substituting equations (4b) and (4d) in the result yield the bulk wave equation

$$\left[\frac{1}{v^2}\partial_t^2 - \Delta + \mu^2\right] M_x^A(\vec{r},t) = 0 \tag{5}$$

with $v = \sqrt{\frac{3}{2}}\lambda J M a$ and $\mu = \sqrt{\frac{6(\gamma^2-1)}{a^2}}$. Identical wave equations hold for the components $M_y^A$, $M_x^B$, and $M_y^B$.

Equation (5) is identical to the wave equation derived by Huang et al. in [30] using the Holstein-Primakov formalism in the linear spin wave approximation and assuming smooth-varying sublattice wave functions; in their work the authors refer to this equation as a relativistic Klein-Gordon equation, which seems inappropriate since equation (5) is not invariant under the Lorentz transformation. The wave equation is hence non-relativistic and can be derived in an identical manner for any three-dimensional lattice with an ordered crystallographic structure (SC, BCC, or FCC structure).

To simplify the algebra, it is useful to define two new variables, namely $\mathcal{M}^A = M_x^A + iM_y^A$ and $\mathcal{M}^B = M_x^B + iM_y^B$. These satisfy the new but equivalent equations of motion of the form

$$-i\partial_t \mathcal{M}^A = 3\gamma\lambda J M \, \mathcal{M}^A + \lambda J M \left(3 + \frac{a^2}{4}\Delta\right)\mathcal{M}^B \tag{6a}$$

$$i\partial_t \mathcal{M}^B = 3\gamma\lambda J M \, \mathcal{M}^B + \lambda J M \left(3 + \frac{a^2}{4}\Delta\right)\mathcal{M}^A \tag{6b}$$

They also satisfy the bulk wave equation (5). Equations (6) are the classical field version of the coupled Harper equations derived by a first order quantum formulation [28-31, 33].

The well-established approach [1 - 8] to solve equation (5), or equivalently equations (6), in a bounded system is to write $\mathcal{M}^A$ and $\mathcal{M}^B$ in the form of a linear combination of the phase factor functions $e^{i(\omega t - k_y y)}e^{q x}$ and $e^{i(\omega t - k_y y)}e^{-q x}$ as follows

$$\mathcal{M}^A = A_1 e^{i(\omega t - k_y y)}e^{q x} + A_2 e^{i(\omega t - k_y y)}e^{-q x} \tag{7a}$$

$$\mathcal{M}^B = B_1 e^{i(\omega t - k_y y)}e^{q x} + B_2 e^{i(\omega t - k_y y)}e^{-q x} \tag{7b}$$

Compared to previous studies [29-31, 33], we here adopt a more general form for the solutions, suitable for bounded systems. These phase factors along the x-axis are intrinsic to the phase field matching theory [16-21]. Note, however, that $e^{qx}$ increases exponentially and is theoretically unphysical in semi-infinite systems where $x \to \infty$. Here, $k_y$ is the continuous wave vector along the infinite y-direction. The real and imaginary values of $q$ correspond respectively to *so-called* evanescent (edge), and propagating (bulk), spin waves in the x-direction along which the nanoribbon is finite. Substituting this linear combination in the bulk wave equation (5) yields the general dispersion relation

$$-\Omega^2 + \frac{3}{2}a^2(k_y^2 - q^2) + 9(\gamma^2 - 1) = 0 \tag{8}$$

with the normalized frequency $\Omega$ defined as $\Omega = \frac{\omega}{\lambda JM}$. This equation (8) should be solved in consistency with the boundary conditions to determine the bulk (propagating) and edge (evanescent) spin wave modes; we note that this equation sets bounds on the real values of $q$ given by $|q| \leq \sqrt{k_y^2 + 6(\gamma^2 - 1)}$.

## *Characteristic boundary equations*

The boundary conditions are derived from the requirement that the edge spins must satisfy the bulk equations of motion [8-11]. Consequently,

$$\partial_t \vec{M}_e^{A/B} = \lambda\, \vec{M}_e^{A/B} \times \vec{H}_e^{A/B} = \lambda\, \vec{M}_e^{A/B} \times \vec{H}_b^{A/B}$$

from which one deduces the effective boundary conditions

$$\vec{M}_e^{A/B} \times (\vec{H}_b^{A/B} - \vec{H}_e^{A/B}) = 0 \tag{9}$$

where $e$ and $b$ respectively stand for edge and boundary. The edge effective fields are obtained using the same approach described above for the bulk fields. The edge fields are indeed different from the bulk fields as the number of nearest neighbors for an edge spin (two for armchair edges) is less than that of a bulk spin.

Equation (9) for the two types of edge spins on the two armchair edge boundaries at $(x = d)$ and $(x = -d)$, yield four boundary conditions for $\mathcal{M}^A$ and $\mathcal{M}^B$ as follows

$$\gamma \mathcal{M}^A(d, y, t) + \mathcal{M}^B\left(d + \frac{a}{2}, y - \frac{a}{2\sqrt{3}}, t\right) = 0 \tag{10a}$$

$$\gamma \mathcal{M}^B(d, y, t) + \mathcal{M}^A\left(d + \frac{a}{2}, y + \frac{a}{2\sqrt{3}}, t\right) = 0 \tag{10b}$$

$$\gamma \mathcal{M}^A(-d, y, t) + \mathcal{M}^B\left(-d - \frac{a}{2}, y - \frac{a}{2\sqrt{3}}, t\right) = 0 \tag{10c}$$

$$\gamma \mathcal{M}^B(-d, y, t) + \mathcal{M}^A\left(-d - \frac{a}{2}, y + \frac{a}{2\sqrt{3}}, t\right) = 0 \tag{10d}$$

With the same expansion technique used in equation (2), equations (10) yield

$$\gamma \mathcal{M}^A(d, y, t) + \left(1 + \frac{a}{2}\partial_x + \frac{a^2}{8}\partial_x^2 - \frac{a}{2\sqrt{3}}\partial_y + \frac{a^2}{24}\partial_y^2\right)\mathcal{M}^B(d, y, t) = 0 \tag{11a}$$

$$\gamma \mathcal{M}^B(d, y, t) + \left(1 + \frac{a}{2}\partial_x + \frac{a^2}{8}\partial_x^2 + \frac{a}{2\sqrt{3}}\partial_y + \frac{a^2}{24}\partial_y^2\right)\mathcal{M}^A(d, y, t) = 0 \tag{11b}$$

$$\gamma \mathcal{M}^A(-d, y, t) + \left(1 - \frac{a}{2}\partial_x + \frac{a^2}{8}\partial_x^2 - \frac{a}{2\sqrt{3}}\partial_y + \frac{a^2}{24}\partial_y^2\right)\mathcal{M}^B(-d, y, t) = 0 \tag{11c}$$

$$\gamma \mathcal{M}^B(-d, y, t) + \left(1 - \frac{a}{2}\partial_x + \frac{a^2}{8}\partial_x^2 + \frac{a}{2\sqrt{3}}\partial_y + \frac{a^2}{24}\partial_y^2\right)\mathcal{M}^A(-d, y, t) = 0 \tag{11d}$$

Substituting equations (8) in the four boundary equations (11) yields a system of four linear equations for the coefficients $A_1$, $A_2$, $B_1$, and $B_2$ which can be written in matrix form as follows

$$\begin{pmatrix} e^{dq}\gamma & e^{-dq}\gamma & e^{dq}t_{+,+} & e^{-dq}t_{+,-} \\ e^{dq}t_{-,+} & e^{-dq}t_{-,-} & e^{dq}\gamma & e^{-dq}\gamma \\ e^{-dq}\gamma & e^{dq}\gamma & e^{-dq}t_{+,-} & e^{dq}t_{+,+} \\ e^{-dq}t_{-,-} & e^{dq}t_{-,+} & e^{-dq}\gamma & e^{dq}\gamma \end{pmatrix} \begin{pmatrix} A_1 \\ A_2 \\ B_1 \\ B_2 \end{pmatrix} = 0 \tag{12}$$

$$t_{\pm,\pm} = \frac{1}{24}\left(24 + a\left(\pm 4i\sqrt{3}k_y - ak_y^2 + 3q(\pm 4 + aq)\right)\right)$$

The determinant of the above $4 \times 4$ matrix should vanish as a necessary condition for the existence of non-zero solutions. Equivalently, at least one of its eigenvalues should be zero. To ensure consistency in the developed theory, the eigenvalues are calculated keeping only linear and quadratic terms in $k_y$ and $q$ for spin waves of long wavelengths. Also, the requirement that one of its eigenvalues vanishes yields the characteristic equations of the form

$$f_1(q, d, \tilde{\gamma}) = 4aq\,Cosh(dq) + (8 + a^2q^2 - 8\tilde{\gamma})Sinh(dq) = 0 \tag{13a}$$
$$f_2(q, d, \tilde{\gamma}) = (8 + a^2q^2 + 8\tilde{\gamma})Cosh(dq) + 4aq\,Sinh(dq) = 0 \tag{13b}$$
$$f_3(q, d, \tilde{\gamma}) = 2aq(-\tilde{\gamma} + Cosh(2dq)) + (2 + a^2q^2 - 2\tilde{\gamma}^2)Sinh(2dq) = 0 \tag{13c}$$

In the long wavelength limit for spin waves, the characteristic functions $\{f_i, i = 1, 2, 3\}$ are hence independent of $\Omega$ and $k_y$, and depend only on $q$, $d$, and $\gamma$. Any of such spin waves hosted by the bounded honeycomb nanoribbon should fulfill consequently one of the $f_i(q) = 0$ equations. Real and imaginary solutions correspond to evanescent and bulk exchange spin wave modes respectively. Note that in our approach, the boundary conditions are solved simultaneously on both edges, unlike previous approaches [29-31, 33] where the boundary conditions are solved separately on the edges, similar to the semi-infinite lattice case.

For the imaginary solutions, it is convenient to substitute $q = ik_x$ ($k_x$ is the wavevector component along the x-direction) in equations (12) and get the equivalent equations

$$g_1(k_x, d, \tilde{\gamma}) = -4ak_x Cos(dk_x) + (-8 + a^2 k_x^2 + 8\tilde{\gamma}) Sin(dk_x) = 0 \tag{14a}$$

$$g_2(k_x, d, \tilde{\gamma}) = \left(2 - \frac{1}{4}a^2 k_x^2 + 2\tilde{\gamma}\right) Cos(dk_x) - ak_x Sin(dk_x) = 0 \tag{14b}$$

$$g_3(k_x, d, \tilde{\gamma}) = 2ak_x(\tilde{\gamma} - Cos(2dk_x)) + (-2 + a^2 k_x^2 + 2\tilde{\gamma}^2) Sin(2dk_x) = 0 \tag{14c}$$

It can be easily shown that none of equations (14) admit continuous solutions in $k_x$ for finite $\tilde{\gamma}$ and $d$. The wavevector $k_x$ along the nanoribbon bounded width $2d$ is hence discrete and the number of solutions depends on this width.

## Numerical Results

For a perfectly isotropic nanoribbon ($\tilde{\gamma} = \gamma = \gamma_e = 1$), equations (13) reduce to

$$q(4Cosh(dq) + aqSinh(dq)) = 0$$

$$(4 + \frac{a^2 q^2}{4})Cosh[dq] + aqSinh[dq] = 0$$

$$qSinh(dq)(aqCosh(dq) + 2Sinh(dq)) = 0$$

None of these equations admit a non-zero real solution, and hence the isotropic magnetic honeycomb nanoribbon does not excite any evanescent modes. For the allowed bulk modes, equations (14) give for the isotropic nanoribbon

$$k_x(-4Cos(dk_x) + ak_x Sin(dk_x)) = 0$$

$$\left(4 - \frac{a^2 k_x^2}{4}\right) Cos(dk_x) - ak_x Sin(dk_x) = 0$$

$$k_x(2 - 2Cos(2dk_x) + ak_x Sin(2dk_x)) = 0$$

*Isotropic wide nanoribbons*

As an example, consider a relatively wide isotropic nanoribbon with $d = 20$ (the lattice constant is set to 1). In the long wavelength part of the Brillouin zone, $|k_x| \leq 0.3$, the boundary characteristic equations yield nine solutions, corresponding to the discrete set of wavevectros $k_x \approx \pm 0.2327$, $\approx \pm 0.1571$, $\approx \pm 0.1532$, $\approx \pm 0.0776$ and $0$. In view of equations (7), the solutions with identical absolute values but of opposite signs can be considered as belonging to the same spin wave mode. In the present case this gives a total of five bulk modes with distinct energy dispersion curves.

Equation (8) can now be used to calculate the normalized energy dispersion curves for the allowed propagating exchange spin waves. These are plotted in figure 2 as a function of the continuous $k_y$ wavevectors for the discrete values of the $k_x$ solutions. The distinct dispersion curves are duplicated at positive and negative values of $k_x$ following the conventional presentation of the dispersion curves for infinite systems. It is worth noting that in the case of a nanoribbon with finite width, the Dirac cone reduces to a single linear dispersion curve as a direct consequence of the bulk spin waves discretization.

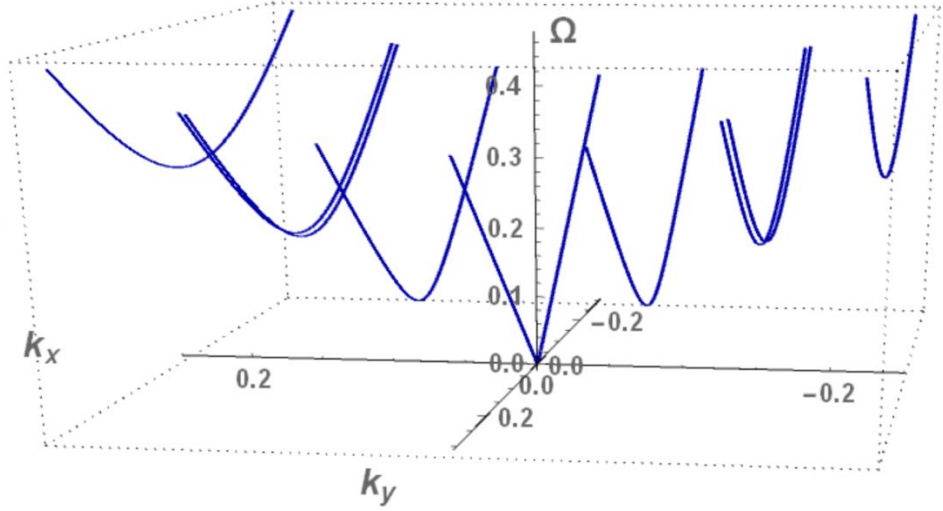

**Figure 2:** Normalized energy dispersion curves $\Omega$ for the discretized bulk exchange spin wave modes on a magnetically isotropic nanoribbon of thickness $2d = 40$, in the long wavelength ($|k_x| \leq 0.3$) strip of the Brillouin zone. They are functions of the continuous $k_y$ wavevector for the allowed $k_x$ discrete values. Otherwise, the isotropic nanoribbon does not permit edge spin wave modes.

The degeneracy of these bulk spin wave modes can be determined from the algebraic multiplicity of the eigenvalues of the linear system presented in equation (11). This analysis proves that the solutions $|k_x| \approx 0.2327$ and $\approx 0.0776$ are degenerate with algebraic multiplicity 2; they constitute double roots for the system's determinant. Moreover, the eigenvector analysis of equation (11) proves that the amplitudes $M_x^A$ and $M_x^B$ are out-of-phase for these modes.

*Anisotropic wide nanoribbons*

This section shows how introducing the least anisotropy gives rise to evanescent modes and lifts the degeneracy for bulk modes. We consider a nanoribbon of width $2d = 40$, characterized by a very small magnetic anisotropy identical on all sites, $\gamma = 1.01$. For the present case, the allowed discrete wavevectors for long wavelengths $k_x$ ($|k_x| \leq 0.3$) obtained by solving equations (14) for bulk spin wave modes, are $k_x \approx \pm 0.2327, \approx \pm 0.2284, \approx \pm 0.1557, \approx \pm 0.1481, \approx \pm 0.0776, \approx \pm 0.0623$ and $k_x = 0$. The degeneracy is lifted completely even with this slight anisotropy, and the number of dispersion curves increases fourfold (2 for positive $k_x$ and 2 for negative $k_x$).

In addition to the allowed bulk modes, equations (12) yield two solutions for edge spin wave modes in the present case, namely for $q \approx \pm 0.0336$, which values correspond to a single evanescent mode. The decay factors for the evanescent mode are very small because of the slight anisotropy. The *bulk* (blue) and *evanescent* (red) dispersion curves for exchange spin wave modes in the large wavelength region are plotted in figure 3. The bulk and evanescent modes are plotted as functions of the continuous $k_y$ wavevector for the allowed discrete $(k_x, q)$ values, $k_x$ for (*bulk*) and $q$ for (*evanescent*) modes, respectively. We present 13 bulk dispersion curves that come in couples at positive and negative non-zero $k_x$ and a single zero $k_x$, which hence correspond to 7 distinct bulk modes, following equations (7). Similarly, the dispersion curves for the evanescent mode are duplicated at $q \approx \pm 0.0336$. The energy of the evanescent mode is found to be very close to that of the bulk mode with $k_x = 0$ because of the very small decay factor.

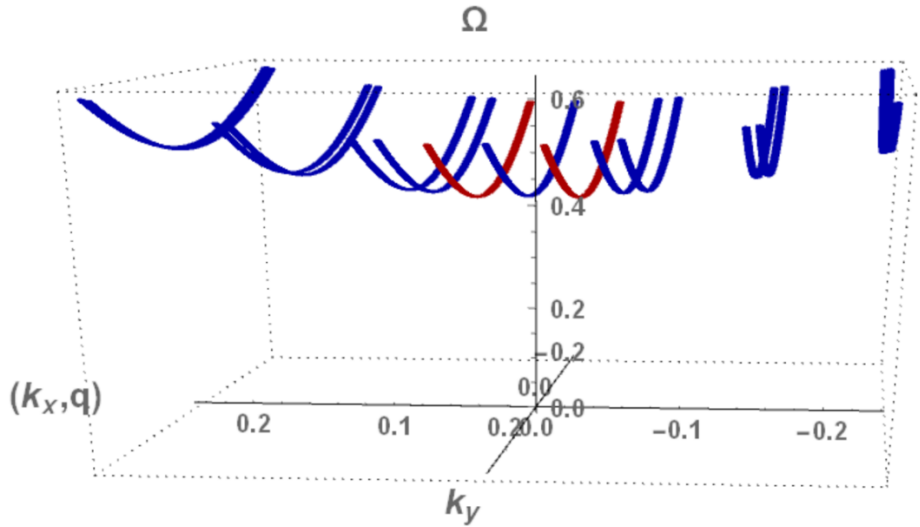

**Figure 3:** Normalized energy dispersion curves $\Omega$ for the discretized *bulk* (blue) and *evanescent* (red) exchange spin wave modes on a magnetically anisotropic nanoribbon of thickness $2d = 40$, with weak anisotropy $\gamma = 1.01$, in the large wavelength ($|k_x| \leq 0.3$) strip of the Brillouin zone. These curves are functions of the continuous $k_y$ wavevector for the allowed discrete $(k_x, q)$ values, $k_x$ (*bulk*) and $q$ (*evanescent*).

The $x$ spatial variation of the amplitudes $M_x^A$ and $M_x^B$ for the evanescent exchange spin wave mode across the finite width of the nanoribbon, $x = [-20, 20]$, can be determined from the eigenvectors of the linear system presented in equation (11). The normalized amplitudes are plotted in figure 4 for $x = [-20, 20]$. As expected from the $q$ values, this mode traverses the nanoribbon and is not confined to the edges. Moreover, the A and B sublattice amplitudes are found to be out-of-phase similar to the bulk modes.

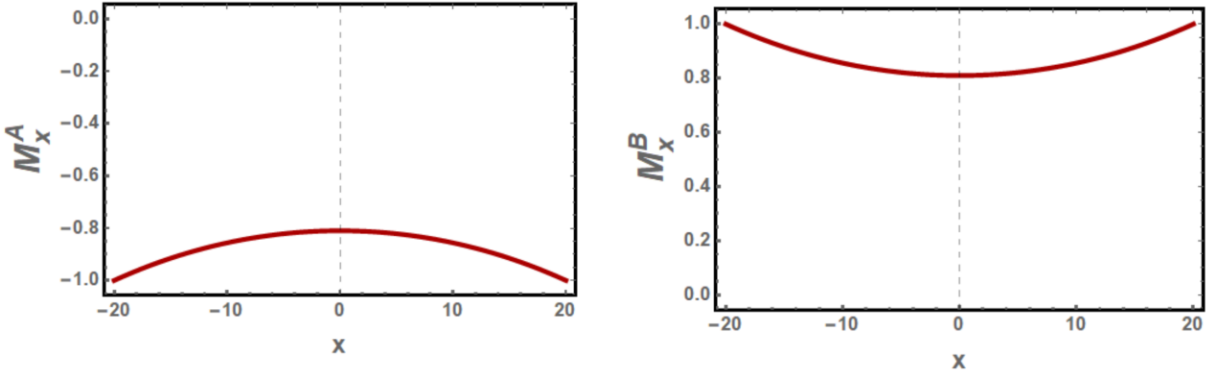

**Figure 4:** The normalized spatial variation of the evanescent spin wave amplitudes along the finite width $2d = 40$ of a weakly anisotropic nanoribbon with $\gamma = 1.01$. The evanescent spin wave traverses the nanoribbon and is not confined to the edges because of the weak anisotropy.

To further analyze the effect of greater magnetic anisotropy on the number and characteristics of the propagating and evanescent exchange spin waves, we next consider a nanoribbon case with $\gamma = 1.1$ keeping the same width $2d = 40$. The allowed discrete wave vectors in the present case, for $|k_x| \leq 0.3$, are $k_x \approx \pm 0.2809$, $\approx \pm 0.2328$, $\approx \pm 0.1945$, $\approx \pm 0.1553$, $k_x \approx \pm 0.1018$, $k_x \approx \pm 0.0776$ and $k_x = 0$. This nanoribbon allows two effectively degenerate evanescent exchange spin wave modes with $q \approx \pm 0.1918$ and $q \approx \pm 0.1907$. The corresponding bulk and evanescent modes are plotted in figure 5. The evanescent modes (red) are effectively degenerate due to the slight difference in the $q$ factors; they are observed to have significantly lower energy than bulk modes (blue) because of the relatively large values of the $q$ factors, which is a direct consequence of the greater anisotropy for this nanoribbon. We also note the important splitting between the bulk modes along the $k_x$ axis induced by the magnetic anisotropy.

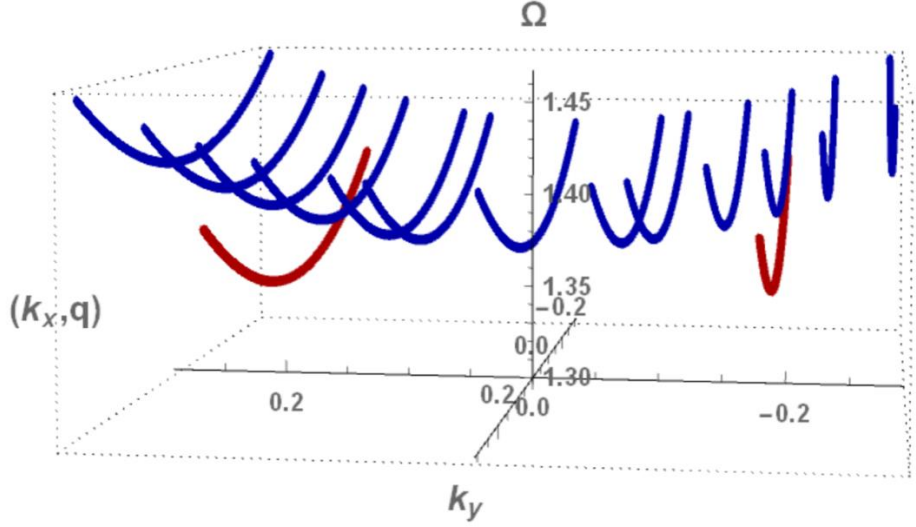

**Figure 5:** As in figure 4 but with a greater nanoribbon anisotropy $\gamma = 1.1$.

The normalized spatial variation of the evanescent spin waves amplitudes across the finite width of the nanoribbon are plotted in figure 6.

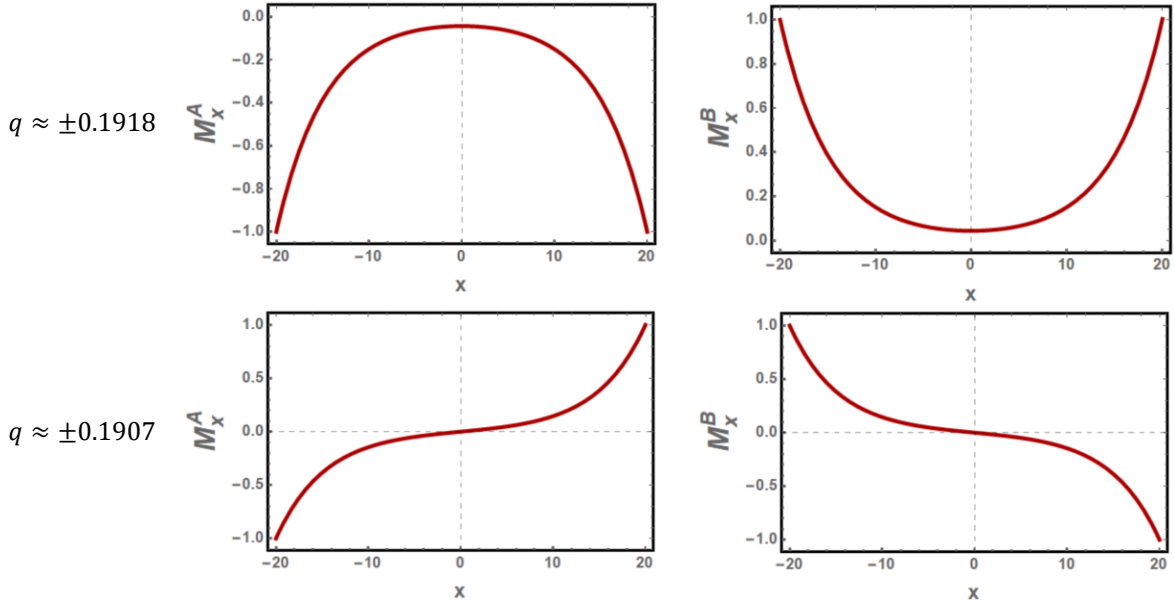

$q \approx \pm 0.1918$

$q \approx \pm 0.1907$

**Figure 6:** The normalized spatial variation of the evanescent spin wave amplitudes across the finite width of a nanoribbon with the augmented anisotropy $\gamma = 1.1$, and $d = 20$. The evanescent spin waves are edge spin waves as a direct consequence of the large anisotropy.

For both modes, the spins on A and B sublattices oscillate out-of-phase. Moreover, the amplitudes decay fast as the spin wave propagates into the bulk of the nanoribbon. The evanescent spin waves are hence highly confined to the edges and can be considered as genuine edge spin waves. These results reveal the importance of magnetic anisotropy for the excitation of edge exchange spin waves.

*Anisotropic narrow nanoribbons*

This edge confinement of the evanescent exchange spin wave modes is lost if the width of the nanoribbon is reduced. To demonstrate this fact, we next consider a narrower nanoribbon with $d = 8$, and $\gamma = 1.1$. The bulk (blue) and evanescent (red) normalized energy dispersion curves are plotted in figure 7. The 2 evanescent modes have factors $q \approx \pm 0.2053$ and $q \approx \pm 0.1685$ and are no more degenerate. Also, the bulk allows only 2 distinct propagating modes with $k_x \approx \pm 0.1907$ and $k_x = 0$.

We note that the non-zero wavelength is very close to the $k_x \approx \pm 0.1945$ solution in the thicker nanoribbon with $d = 20$. Nevertheless, the evanescent spin waves penetration length into the bulk in this case is relatively larger than for wide nanoribbons as shown in figure 8.

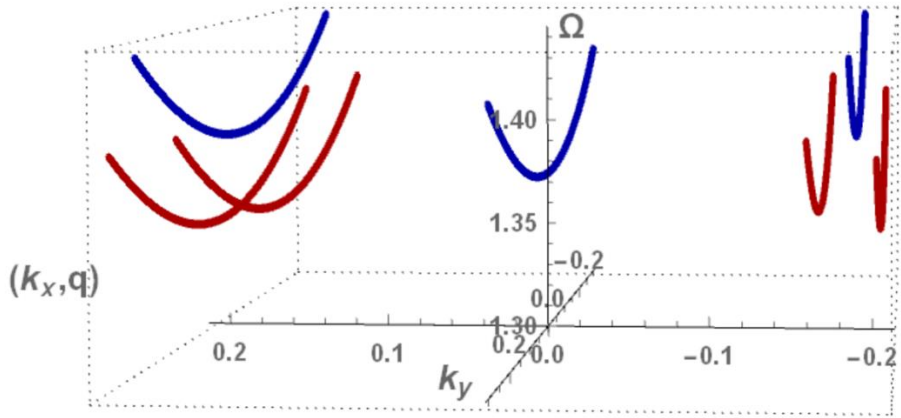

**Figure 7:** As in figure 5 but with a narrow nanoribbon $d = 8$.

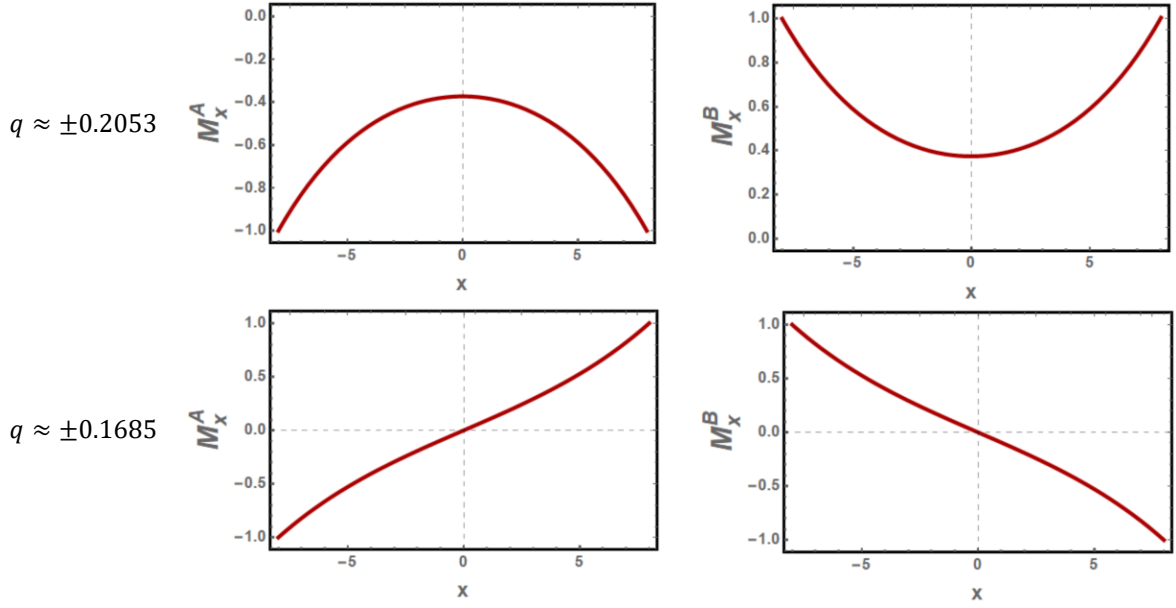

**Figure 8:** The normalized spatial variation of the evanescent spin waves amplitudes along the finite width of a narrow nanoribbon with significant anisotropy $\gamma = 1.1$ and $d = 8$. The evanescent spin waves penetrate significantly into the bulk.

*Conclusion*

We systematically developed a spin wave field theory to study the long wavelength exchange spin wave excitations on honeycomb nanoribbons characterized by the Néel antiferromagnetic order and armchair edge boundaries. The formalism is based on the semi-classical Heisenberg Hamiltonian and the classical field equations of motion for the spins. Boundary conditions are established, requiring that the edge spins satisfy the bulk equations of motion. A set of characteristic boundary equations, dependent solely on the finite width of the nanoribbon and the exchange anisotropy parameters, is hence formally derived. To determine the spin wave excitations on these nanoribbons, the Bloch equations of motion are solved consistently with the derived characteristic boundary equations. Compared to previous studies, our approach assumes a more general form for the magnetization dynamics in the nanoribbons. Furthermore, the boundary conditions are derived and solved simultaneously on both edges.

Our study highlights the important effect on bulk and edge spin waves induced by the finite width of the nanoribbon and the magnetic exchange anisotropy. We predict the discretization of the bulk spin wave spectrum as a direct consequence of the finite width of the nanoribbons. The allowed discrete wavevectors along the finite width of the nanoribbons depend on the width itself and the nanoribbon magnetic anisotropy. Further, this latter is shown to lift the degeneracy of the allowed discrete bulk modes. With the discretization of the bulk modes, the Dirac cone for infinite (or semi-infinite) magnetically isotropic honeycomb lattice is shown to reduce to a single linear dispersion curve for the nanoribbon lattice.

The magnetic anisotropy is found to be a necessary condition for the existence of evanescent modes with non-linear dispersion for the nanoribbons with armchair edge boundaries. The spatial variation of the spin waves amplitudes across the nanoribbon widths is determined for the evanescent modes. Genuine edge modes in these nanoribbons are predicted to exist for only relatively wide nanoribbons and significant anisotropy.

Our developed theoretical approach is general, and can be applied for nanoribbons with arbitrary edge types and any possible magnetic ordering of the honeycomb lattice spins. We have already tested the method on zigzag boundary edges, obtaining results identical to those based on quantum model approaches for the isotropic case. Our spin wave field theory results, obtained for honeycomb nanoribbons with (zigzag, zigzag), (bearded, bearded) and (zigzag, bearded) boundary edges, will be presented elsewhere.